\documentstyle[prl,aps]{revtex}

\begin{document}
\twocolumn
%\draft
\title{Vortex coupler for atomic Bose-Einstein condensates}
\author{Karl-Peter Marzlin and Weiping Zhang}
%\\[2mm]
\address{
School of Mathematics, Physics,
Computing and Electronics, Macquarie University, Sydney, NSW 2109,
Australia}
\author{Ewan M. Wright}
\address{Optical Sciences Center, University of Arizona, Tucson,
	Arizona 85721}
\maketitle
%%%%%%%%%%%%%%%%%%%%%%%%%%%%%%%%%%%%%%%%%%%%%%%%%%%%%%%%%%%%%%%%
\begin{abstract}
A coherent coupler is proposed to spin a Bose-Einstein
condensate composed of ultracold alkali atoms into a vortex state
(VS). 
The proposal is based on a Raman transition induced by two 
copropagating $\sigma^+$ and $\sigma^-$ polarized Laguerre-Gaussian
laser beams with different frequencies. We show that the 
transfer of angular momentum of photons to the condensed atoms through
a Raman transition leads to a coherent coupling of the ground-state 
condensate to a rotating condensate in a VS. The detection of
such a VS is discussed.
\end{abstract}
%%%%%%%%%%%%%%%%%%%%%%%%%%%%%%%%%%%%%%%%%%%%%%%%%%%%%%%%%%%%%%%%
%\narrowtext
It is well-known that VS's play a central role in characterizing
the superfluid properties of large-size Bose systems such as superfluid
helium (see, e.g., ef. \cite{rotation1}).
Recently the experimental realizations of 
Bose-Einstein condensation in trapped ultracold alkali atomic gases
\cite{experimente} have generated great interest in studying the superfluid
aspects of the small-scale trapped Bose gases \cite{rotation2}. To understand
how the small-scale Bose-Einstein condensate is related to a superfluid, 
it is natural to study the rotational properties of the condensate 
and to examine VS's.
Being different from superfluid helium, the trapped Bose gases 
are not in direct contact to an external container.
How to rotate such gases and to create VS's is still an open question.

In this letter, we propose to employ
a vortex coupler to realize such a goal. The principle for the vortex coupler
is illustrated in Fig.1. Two copropagting $\sigma^+$ and $\sigma^-$ 
polarized laser beams along the z direction (gravity's direction) are used to 
induce a Raman transition between two hyperfine levels of
ground-state alkali atoms. Initially we assume that a Bose-Einstein 
condensate is prepared in the trapping state $|- \rangle =|F=1,M_F=-1\rangle$. 
We consider the time evolution of the condensate after the trap is switched 
off and the Raman laser beams are applied. In addition, gravity 
effects are excluded here since we are only interested in the rotational  
motion of the condensates in the x-y plane. To avoid the destructive 
incoherent heating of the condensate due to spontaneous decay of excited 
states $|j \rangle$, the two Raman beams are detuned by a frequency $\Delta$
from the optical transition between the ground state and the excited state
manifolds. In this case adiabatic 
elimination of the excited states $|j \rangle$ results in a nonlinear
Schr\"odinger equation describing the coherent Raman-type coupling \cite{Zeng}  
between the condensate wave functions $\phi_-$ and $\phi_+$ corresponding to 
the magnetic sublevel $|-\rangle$ and $|+\rangle=|F=1,M_F=1\rangle$: 
\begin{eqnarray}
  i\hbar \dot{ \phi}_- &=& ( T -\mu ) \phi_-
  + \frac{4\pi\hbar^2 a_{sc} N}{m} \{2 |\phi_-|^2 + |\phi_+|^2 \} \phi_-
  \nonumber \\ & &
  +\frac{\hbar |\Omega_1^{(-)}|^2}{4 \Delta } 
  \phi_-  + \frac{\hbar \Omega_1^{(-)} \Omega _2^{(+)}}{4 \Delta }
  e^{-i \Delta \omega t} \phi_+
  \nonumber 
\\
  i\hbar \dot{ \phi}_+ &=& ( T -\mu ) \phi_+
  + \frac{4\pi\hbar^2 a_{sc} N}{m} \{2 |\phi_+|^2 + |\phi_-|^2 \} \phi_+
  \nonumber \\ & &
  +\frac{\hbar |\Omega_2^{(-)}|^2}{4 \Delta } 
  \phi_+  + \frac{\hbar \Omega_1^{(+)} \Omega _2^{(-)}}{4 \Delta }
  e^{i \Delta \omega t} \phi_-
\label{1stdgl} \end{eqnarray}
where $\Omega_i^{(+)}$ denote the positive frequency part of the corresponding
Rabi frequencies of the two Raman beams. $\Delta \omega := \omega_2-\omega_1$
is the frequency difference between the two Raman beams, $m$ the atomic 
mass, $a_{sc}$ the scattering length, and $N$ the total number of atoms 
in the initial condensate. $\mu$ denotes the chemical potential needed to 
fix the mean number of atoms. We have normalized the condensate wave functions
by the condition $\int d^3 x \{|\phi_+|^2 + |\phi_-|^2\} =1$. For the sake 
of a concise notation we have defined the operator $T:= \vec{p}^2/(2m)$
for the kinetic energy .

The last terms in Eqs. (\ref{1stdgl})
represent the Raman coupling which coherently
transfers the initial condensate in the level $|- \rangle$ to 
the initially empty level $|+ \rangle$. Evidently, if the Raman 
laser beams are generated by ordinary laser sources with Hermite-Gaussian modes, 
Eqs. (\ref{1stdgl}) only describe a normal coherent coupler which doesn't
cause a rotation of the condensate since no angular
momentum is transferred by the coupling term. Thus, to realize 
a coherent rotational coupler, the Raman laser beams must carry an
angular momentum. In view of the recent experimental progress in laser
physics and atom optics \cite{angular2}, a natural choice is to employ 
Laguerre-Gaussian laser beams. For this kind of modes
the Rabi frequency $\Omega _1^{(+)}$ for the
$\sigma ^+$ polarized beam given by \cite{angular2}
\begin{equation} \Omega_1^{(+)}(\vec{x}) = \Omega_0 e^{-r^2/w^2}
  \left ( \frac{\sqrt{2}r}{w}\right )^l e^{il\varphi}e^{ikz}
\label{lag} \end{equation}
($l\geq 1$), where $r^2 := x^2 + y^2$. A similar expression applies to the
$\sigma ^-$ polarized beam $\Omega_2^{(+)}$ with $\varphi$ replaced by
$-\varphi$. By analogy between quantum mechanics and paraxial optics,
the azimuthal angular dependence of the phase factor in Eq. (\ref{lag})
determines an orbital angular momentum of $l \hbar$ per photon in the
Raman beams. Here we consider the simple case that both Raman beams
are in the first Laguerre-Gaussian
mode ($l=1$) and their waist $w$ is much larger
than the size of the initial condensate. We then can approximate 
the above expression to $\Omega_1^{(+)} =
\sqrt{2} \Omega_0 (x+iy)/w \exp (ikz)$
and likewise for $\Omega _2^{(+)}$ with $x+iy$ replaced by $x-iy$.
Inserting these expressions into Eq. (\ref{1stdgl})
we obtain
\begin{eqnarray}
  i\hbar \dot{\phi}_- &=& (T+V -\mu) \phi_- + V e^{-2i \varphi}
  e^{-i \Delta \omega t} \phi_+ \nonumber \\ & &
  + \frac{4\pi\hbar^2 a_{sc} N}{m} [2|\phi_-|^2 +|\phi_+|^2]\phi_-
  \label{dgl2a} \\
  i\hbar \dot{\phi}_+ &=& (T+V -\mu) \phi_+ + V e^{2i \varphi}
  e^{i \Delta \omega t} \phi_- \nonumber \\ & &
  + \frac{4\pi\hbar^2 a_{sc} N}{m} [2|\phi_+|^2 +|\phi_-|^2]\phi_+ \; .
\label{dgl2b} \end{eqnarray}
We see that, because of the donought-shaped transverse profile of 
the Laguerre-Gaussian mode, the light-induced potentials and 
the Raman coupling terms have approximately the spatial structure 
of a harmonic oscillator potential 
$V:=(1/2) m \omega_{\mbox{{\scriptsize eff}}}^2 r^2$, where the effective 
oscillator frequency is defined as $\omega_{\mbox{{\scriptsize eff}}} :=
\sqrt{\hbar/(m \Delta )} |\Omega_0|/w$.
In order to avoid excitations of the condensate after
the original magnetic trap is switched off, we choose the appropriate
laser intensities so that the effective oscillator frequency
$\omega_{\mbox{{\scriptsize eff}}}$
coincides with the transverse oscillating frequency of the original
magnetic trap. In this sense, the donought-shaped Laguerre-Gaussian
mode spatially acts as a waveguide for condensates where the transvere motion  
of the condensates is confined by the light-induced potentials. 
In addition, one more important feature of Eqs.
(\ref{dgl2a}, \ref{dgl2b}) is that the Raman coupling terms
now contain phase factors $\exp (\pm 2i \varphi)$.

The physics implicit  
in Eqs. (\ref{dgl2a}, \ref{dgl2b}) is very obvious. As sketched in 
Fig.1, atoms in the condensate corresponding to level $|-\rangle$ absorb 
photons from the $\sigma^+$ beam and make a Raman-transition to the
$|+\rangle$ level by emitting photons into $\sigma^-$ beam.
Each cycle of photon absorption and emission in the Raman transition 
directly leads to the phase factors which describe
the transfer of an orbital angular momentum $2 \hbar$, originating from two 
photons in the $\sigma^+$ and $\sigma^-$ laser beams, to the condensed atoms.
As a result, the condensate fraction in the
state $|+\rangle $ is set into a VS with twice the elementary quantum
of circulation by the coherent Raman transfer.

To quantitatively estimate the Raman transfer of the non-rotating 
condensate in level $|-\rangle $ to the vortex-state condensate 
in level $|+\rangle$, we further reduce Eqs.
(\ref{dgl2a}, \ref{dgl2b})
by assuming separately for each of the two states
that the spatial part of the condensate varies little during the Raman 
coupling. For the present case this approximation is justified
by two reasons: (1) phonon excitations of the condensate $\phi_-$ are avoided by 
choosing the light-induced potential V to match the original magnetic 
trapping potential, and (2) the spatial dependence of the Raman coupling 
term in Eq. (\ref{dgl2b})
is proportional to $ r ^2 $ which exactly leads to
a transverse spatial structure matching the transverse spatial shape 
required by a VS with two quantum circulations. 

Using this approximation we can make the following ansatz for the
condensate wavefunctions,
$\phi_-(\vec{x},t) = \alpha (t) \exp [i (\mu/\hbar - \kappa)t]$ $
\psi_g(\vec{x})$ and
$\phi_+(\vec{x},t) = \beta (t) \exp [i ( \Delta \omega + \mu/\hbar
- \kappa )t] \psi_v(\vec{x})$, with $\psi_g (\vec{x})$ denoting
the spatial dependence of the nonrotating condensate
and $\psi_v$ representing the VS 
with two elementary quanta of circulation. Further,
we approximate the spatial parts $\psi_g (\vec{x})$ and
$ \psi_v(\vec{x})$ of the wavefunctions by that of a
trapped ideal Bose gas
\cite{note}. The spatial wavefunctions are then given by
$\psi_g = \exp \{-(1/2)[(r/L_\perp)^2 + (z/L_z)^2]\}/(
\pi^{3/4}L_\perp L_z^{1/2})$ for the ground state and
$ \psi_v = (x+iy)^2\psi_g /(\sqrt{2} L_\perp^2)$ for the VS.
Here $L_\perp$ and $L_z$ denote the size parameters of the
original trap in the x-y plane and the z-direction, respectively.
The parameter $\kappa $ determines the strength of the interatomic
interaction and is defined as $\kappa := \pi\hbar a_{sc} N/(m (2\pi)^{3/2}
L_\perp^2 L_z)$.
Projecting Eqs.
(\ref{dgl2a}, \ref{dgl2b}) on these two states we arrive at a
system of two ordinary differential equations,
\begin{eqnarray} 
  i \dot{ \alpha } &=& \frac{ \omega _{\mbox{{\scriptsize eff}}}
  }{\sqrt{2}} \beta + 7\kappa |\alpha |^2 \alpha
  \label{dgl3a} \\
  i \dot{\beta} &=& (\Delta \omega + 2
  \omega_{\mbox{{\scriptsize eff}}} ) \beta +
    \frac{ \omega_{\mbox{{\scriptsize eff}}}
  }{\sqrt{2}} \alpha + 2 \kappa |\beta|^2 \beta \; ,
  \label{dgl3b}
\end{eqnarray}
where $|\alpha |^2$ gives the occupying fraction of the condensate in 
level $|-\rangle$ and $|\beta |^2$ that of the VS in level
$|+\rangle$. Because of probability conservation, we have the constraint
$|\alpha |^2 + |\beta |^2 =1$. The terms proportional to
$\omega_{\mbox{{\scriptsize eff}}}
/\sqrt{2}$ describe the laser-induced Raman coupling between the non-rotating
condensate and the VS. The nonlinear contributions 
proportional to $\kappa $ have their origin in the interatomic interaction.
As the strength
of the nonlinear interaction depends on the density of the condensate, 
which is different for the non-rotating condensate and the VS, 
the nonlinear terms are of different magnitude for the two states.
Finally, the expression linear in $\beta $ of Eq. (\ref{dgl3b})
includes the energy difference $2 \hbar
\omega_{\mbox{{\scriptsize eff}}}$ between
the VS and the ground state as well as the energy transfer
$\hbar \Delta \omega $ from the laser beams to the condensate.

To solve Eqs. (\ref{dgl3a}, \ref{dgl3b}) it is convenient to consider
the population difference $f(t):= |\alpha |^2 - |\beta |^2$ which also
is the physical quantity we are interested in. By using the initial
condition $f(0)=1, \dot{f}(0)=0$ (all atoms are initially in the
$|-\rangle $ state) one can derive from the system of equations
(\ref{dgl3a}, \ref{dgl3b}) the first order equation
\begin{equation} 
  \dot{f}^2 = (1-f) \left \{ 
  2 \omega_{\mbox{{\scriptsize eff}}}^2 (1+f)
  - \frac{\kappa^2}{16} (1-f)
   (19 - \varepsilon + 9f)^2 \right \} \; ,
\label{dgl4}\end{equation}
with $\varepsilon := 4(\Delta \omega + 2
\omega_{\mbox{{\scriptsize eff}}})/ \kappa $.
Eq. (\ref{dgl4}) indicates that $f(t)$ can be expressed in term of
elliptic functions \cite{abramo64}. The exact analytical solution
is given by
\begin{equation} 
  f(t) = \frac{a \; \mbox{dn}(i \nu t | m) + b}{\mbox{dn}(i\nu t|m) + a+b-1} \; .
\label{lsg}\end{equation} 
$\mbox{dn}(z|m)$ is one of the Jacobian elliptic functions. 
The parameter $a$ is determined by one of the real roots of a 6th order
polynomial $P(\xi) = \sum_{i=0}^6 g_i \xi^i$, where $g_i$ are 2nd order
polynomials in the parameters $\varepsilon, \kappa$. The other parameters
$b, m$, and $\nu$ can be expressed as a rational polynomial of 5th
order in the parameter $a$.
The explicit expressions are rather lengthy and will therefore
not be given here. Instead, we will rely on a
numerical calculation to determine $a,b,m,$ and $\nu$.

As an example we use the experimental parameters of the $^{87}$Rb
experiment \cite{jila}. The number $N$ of condensed atoms is
4500, and the laser intensity to be chosen so that the
effective oscillator frequency $\omega_{\mbox{{\scriptsize eff}}}$
is equal to the transversal trap frequency $\omega_\perp
= 132$ Hz. The trap size parameters are given by
$L_\perp = 2.35 \mu$m and $L_z = 1.4 \mu$m. The scattering
length is taken to be $a= 5$nm so that the nonlinear
coupling parameter becomes $\kappa = 422$ s$^{-1}$.
In Fig. 2 we show three solutions corresponding to different values
of the frequency difference $\Delta \omega$. The dotted line
corresponds to $\Delta \omega = 2700$ Hz for which the parameters
in Eq. (\ref{lsg}) take the values $a=0.86$, $b=0.39$, $m=1.07$,
and $\nu = 514$ Hz. As will be explained below the dot-dashed line
($\Delta \omega = 1962$ Hz,
$a=0.72$, $b=0.70$, $m=656$, $\nu=16.4$ Hz) is a very peculiar
case for which the maximal population transfer is reached
(for fixed $\Delta \omega$).
Within our model about one half of the $^{87}$Rb atoms could
be transferred to the $|+\rangle$ state. The dashed curve
shows that for a slightly smaller frequency difference
($\Delta \omega = 1850$ Hz,
$a=0.70$, $b=0.74$, $m=-2.97$, $\nu=236$ Hz) the
solution does only perform relatively small
oscillations.

In order to better understand this
behaviour we consider the equation which determines the
extremal points in the time evolution of the population difference,
$\dot{f}=0$. According to Eq. (\ref{dgl4}) these extremal
points are determined by the (real) roots of a 4th order polynomial
in $f$. This allows us to determine the
maximum population transfer for a given $\Delta \omega$.
In Fig. 3 we show the real roots of the polynomial given by the
r.h.s. of Eq. (\ref{dgl4}) as a function of $\Delta \omega$.
$f=1$ is always a root, and since we start just with this value
(all atoms in the $|-\rangle$ state) the solution can only
oscillate between $f=1$ and the next root for a given
$\Delta \omega$. It can be read off from Fig. 3 that the
maximum population transfer is about 50\% and appears around
$\Delta \omega = 1962$ Hz (solid line in Fig. 2). The shape of
the oscillations for frequency differences around this value
is heavily affected by the fact that the polynomial becomes
very small in the middle of the cycle, causing a large delay
in the time evolution. For slightly smaller values of $\Delta
\omega$ a new root appears relatively close to 1. This is
responsible for the fact that the maximum population transfer
is dramatically reduced if $\Delta \omega$  passes the
critical value around 1950 Hz.

Though a population transfer of 50\% is satisfactory it would be
desirable to transfer almost the whole condensate into the
VS. Fig. 3 indicates that this is only possible if
we can reach the peak at $f=-1$ (corresponding to a
complete population transfer). 
As is shown above this is
impossible for a fixed $\Delta \omega$ if the system starts from $f=1$.
But this restriction
may be circumvented by introducing a linearly time-dependent
frequency difference between the two lasers. We have numerically
solved Eqs. (\ref{dgl3a}, \ref{dgl3b})
by assuming that the frequency difference
varies with $\Delta \omega (t) = 2900$ Hz - $(3336$ Hz $)^2 t$.
Obviously this leads to almost
complete transfer of the condensate to the VS.
This effect can be explained by noting that
the nonlinear interaction between
the atoms causes a time-dependent
energy shift between the two states.
This energy shift is due to the fact that the nonlinear interatomic
interaction is proportional to the populations which change with
time. Thus, after
some time the transition will be out of resonance even if it was
initially at resonance. A phase modulation of $\Delta \omega$
can compensate for this time-dependent energy shift.

So far we have shown how to create a VS by coherent Raman coupling.
How to detect the VS remains to be discussed.
A direct detection of the VS is to observe the
off-resonance absorption image of the rotational cloud. For a vortex
state one should expect a bright "hole" in the image which accounts
for the vortex "core" in the density distribution of the
VS. Another possibility to detect a VS is to
observe the Doppler frequency shift due to the quantized
circular motion of the atoms. Here we focus on discussing the
absorption image for the circular motion.

To realize a non-destructive detection scheme we propose to
employ two counterpropagating weak $\sigma^+$
and $\sigma^-$ probe beams in the $x$-direction and to observe the absorption
image of the $\sigma^-$ probe beam. The principle for the detection is simply
based on Raman absorption of the $\sigma^-$ beam (if the VS is
prepared in the level $|+\rangle$). The Raman absorption coefficient of the
$\sigma^-$ probe beam is proportional to the probability of the Raman
transition from $|+ \rangle$ to $|- \rangle$ which has the form
\cite{unp}
\begin{equation}
 \gamma (y,z) \propto \int_{-\infty}^\infty dx
 \frac{|\Omega_{-}|^2}{W} \sin \big (
  W t_0 \big )
 |\psi_v (\vec{x})|^2
\label{abs} \end{equation}
with $W := \sqrt{\delta^2 + |\Omega_+|^2 |\Omega_-|^2/(4 \Delta^2)} $.
$\Omega_\pm$ are the Rabi frequencies of the two probe-beams.
$\delta = \Delta \omega -2(\hbar k_L^2/m) -\Delta \omega_v$ is the effective
frequency difference in the two probe-beams with $\Delta \omega_v = 4(\hbar
k_L/m) \vec{e}_x\cdot \nabla \varphi$ being the Doppler
frequency shift
induced by circular motion of the atoms in the VS.
Eq. (\ref{abs})
clearly shows that the Raman absorption image directly displays the
rotational properties of the VS. Fig. 4 shows the image of a
VS discussed in this paper.
We see that the absorption is asymmetric around the vortex axis.
Such an asymmetry directly gives signature for the Doppler effect induced
by the quantized circular motion of the atoms in the vortex
state. Since the $\sigma^-$ probe beam propagates along the
$x$-direction, the strong absorption (black spot in Fig. 4)
corresponds to atoms with velocity antiparallel to the
wavevector $k_L \vec{e}_x$ of the $\sigma^-$ beam. The weak
absorption on the opposite side of the vortex line appears because
there the rotating atoms are moving parallel to the $\sigma^-$ beam.
%%%%%%%%%%%%%%%%%%%%%%%%%%%%%%%%%%%%%%%%%%%%%%%%%%%%%%%%%%%

{\bf Acknowledgement}: The work has been supported by the
Australian Research Council.
%%%%%%%%%%%%%%%%%%%%%%%%%%%%%%%%%%%%%%%%%%%%%%%%%%%%%%%%%%%%%%%%
%\newpage

\newpage
{\bf Figure captions:}\\[1cm]
{\bf Fig. 1:} Two Laguerre-Gaussian laser beams induce a
Raman transition and generate a VS in the condensate.
\\[5mm]
{\bf Fig. 2:} The population difference between the two
ground states as a function of time for different constant values
of the frequency difference between the lasers, and for a linearly
varying frequency difference (solid line).
\\[5mm]
{\bf Fig. 3:} The extremal points of the population difference
as a function of $\Delta \omega$ (in Hz). The time evolution starts
at $f=1$ and moves downwards until it hits another extremal point.
\\[5mm]
{\bf Fig. 4:} A typical Raman absorption image of a VS.
The Doppler shift of the rotating atoms leads to a higher
absorption if the atom move towards the $\sigma^-$ laser.
$y$ and $z$ are given in units of the trap size parameter.

\begin{thebibliography}{99}
\bibitem{note} For the realistic case of an interacting Bose gas,
	$\psi_g$ and $\psi_v$ could be obtained
	by using a Gaussian variational ansatz \cite{baym96}.
	This would lead to numerical corrections to the parameters
	in Eqs. (\ref{dgl3a}, \ref{dgl3b}).
	However, such corrections do not affect the conclusions
	and do anly modify the realistic estimation for other
	parameters such as the frequency $\Delta \omega$.
\end{thebibliography}
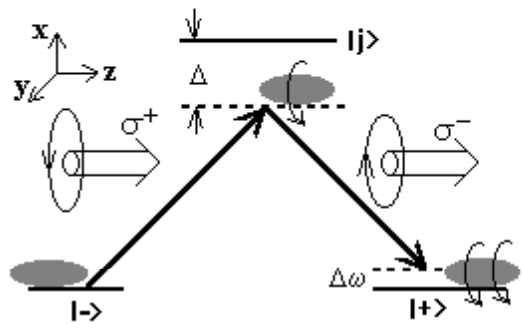
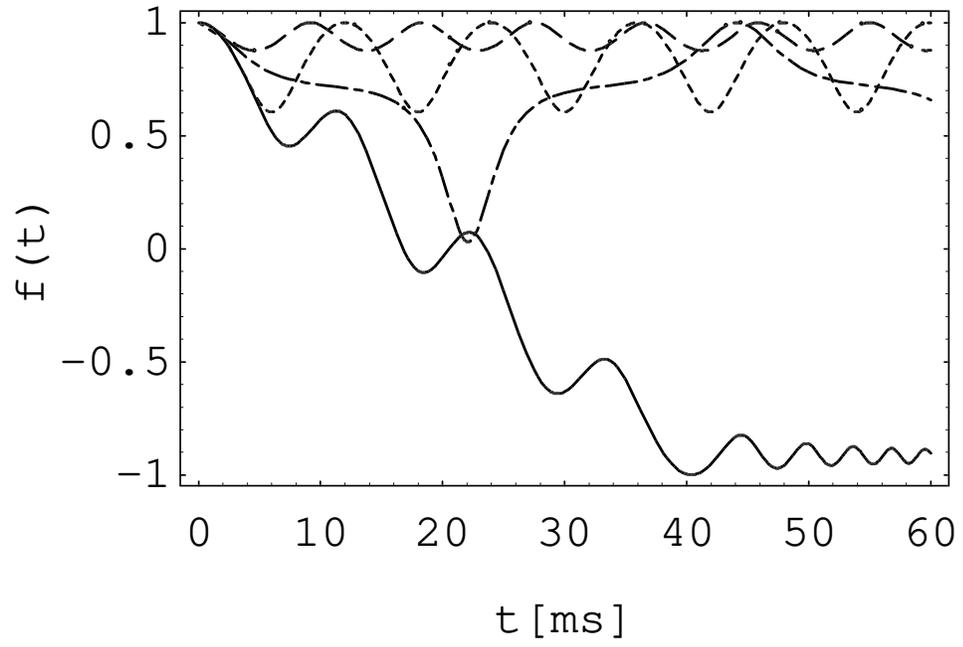
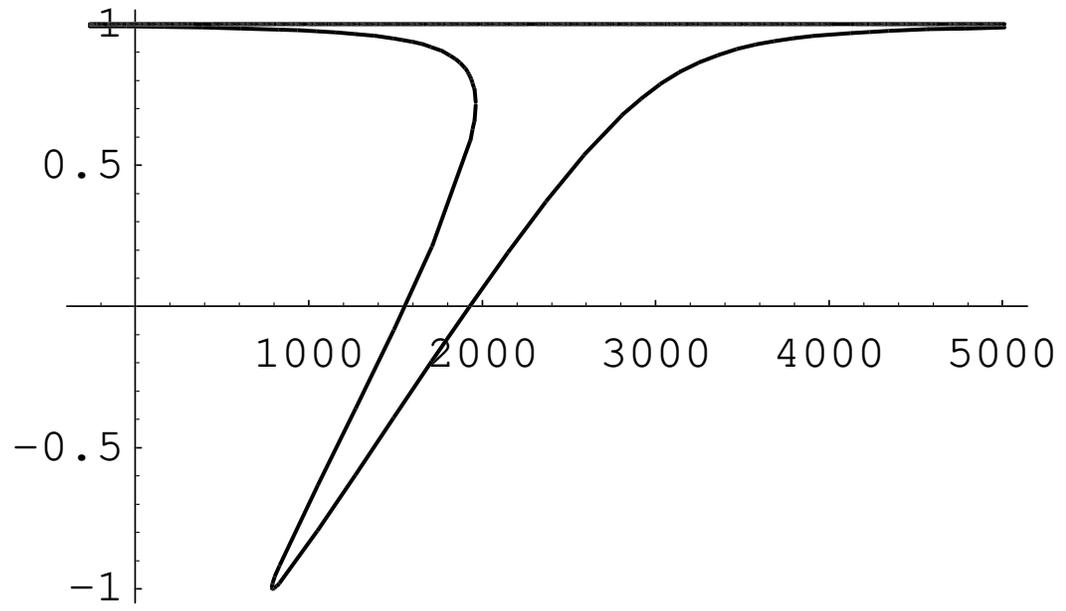
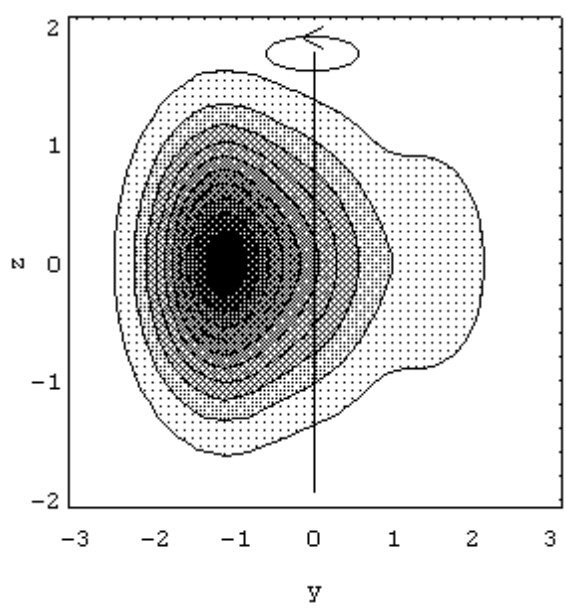
\end{document}